\documentclass[twocolumn,showpacs,preprintnumbers,amsmath,amssymb]{revtex4}
\usepackage{graphicx}% Include figure files
\usepackage{dcolumn}% Align table columns on decimal point
\usepackage{bm}% bold math

\begin{document}

\preprint{Meckler et al.}

\title{Commensurability Effects in Hexagonal Antidot Lattices\\
 with Large Antidot Diameters}% Force line breaks with \\

\author{S. Meckler}\author{T. Heinzel}
\email{thomas.heinzel@uni-duesseldorf.de}
\affiliation{Heinrich-Heine-Universit\"at, Universit\"atsstr.1,
40225 D\"usseldorf, Germany}

\author{A. Cavanna}\author{G. Faini}\author{U. Gennser}\author{D. Mailly}
\affiliation{CNRS-LPN, Route de Nozay, 91960 Marcoussis, France}

\date{\today}

\begin{abstract}
The observation of a novel type of commensurability resonance in
two-dimensional, hexagonal antidot lattices is reported. These
resonances have a classical character and occur at magnetic fields
above the resonance that corresponds to the cyclotron motion
around a single antidot. The resonances are visible only for
antidots with effective diameters larger than 50 \% of the lattice
constant. Simulations reveal that they originate from quasi-stable
electron trajectories that bounce between three neighboring
antidots. This interpretation is backed by the observation of
large-period Aharonov-Bohm type oscillations at low temperatures.
\end{abstract}

\maketitle

\section{\label{sec:1}INTRODUCTION}

The study of electron transport in artificial, two-dimensional
periodic potentials has revealed a variety of interesting
phenomena over the past 15 years. One variant of such systems are
antidot lattices, i.e. periodic potentials with maxima above the
Fermi energy of the two-dimensional electron gas (2DEG)
\cite{Ensslin1990,Weiss1991,Lorke1991}. Most strikingly,
resonances in the longitudinal magnetoresistance are found, which
can be interpreted in terms of classical cyclotron orbits that are
commensurate with the antidot lattice \cite{Weiss1991}. A more
thorough classical treatment based on the Kubo formalism
\cite{Fleischmann1992} has revealed that these resonances actually
have their origin in the magnetic field dependent mixture of
chaotic and regular trajectories, where the latter remain pinned
in weak electric fields. Moreover, this theory provides an
explanation for the observation of a negative Hall effect in weak
magnetic fields $B$ \cite{Fleischmann1994}. Another striking
effect is the occurrence of $B$-periodic oscillations, which have
been explained within a semiclassical theory by a few yet dominant
quantized periodic orbits \cite{Weiss1993,Richter1995}. More
recently, signatures of the famous fractal energy spectrum of such
potentials, also known as the Hofstadter butterfly, have been
observed
\cite{Albrecht2001,Schlosser1996}.\\
A large fraction of these studies was performed on square
\cite{Ensslin1990,Weiss1991,Lorke1991,Eroms1999,Weiss1993,Albrecht2001,Schlosser1996}
or rectangular \cite{Schuster1993} lattices, while only a few
experiments on hexagonal lattices have been published
\cite{Weiss1994,Nihey1995,Ueki2004,Iye2004}. For the majority of
the effects, the lattice type is, in principle, irrelevant.
Nevertheless, hexagonal lattices show some peculiarities that are
absent in other types of antidot lattices. In particular,
Altshuler-Aronov-Spivak oscillations have been observed around
$B=0$ in hexagonal lattices \cite{Nihey1995,Iye2004}, while
Aharonov-Bohm oscillations can be detected at larger magnetic
fields \cite{Ueki2004,Iye2004}. Also, it has been suggested
recently that scattering centers with a short range hexagonal
order may be responsible for a phenomenology resembling that one
observed in metal-insulator transitions in two dimensions
\cite{Heinzel2003}. Moreover, a detailed understanding of the
transport in two-dimensional hexagonal lattices is of broad
relevance, since this type of lattice forms via self-organization
on a mesoscopic scale in a variety of systems, such as vortex
lattices in type II - superconductors \cite{Tinkham2004}
or diblock copolymers \cite{Thurn2000}.\\

Here, we report the observation of a novel type of
magnetotransport resonance in hexagonal antidot lattices with
large effective antidot diameters $d_e\geq 0.5 a$, where $a$
denotes the lattice constant. These resonances occur at cyclotron
radii smaller than the antidot radius. Their signatures are
visible only in the longitudinal component of the
magneto-resistivity tensor, while temperature dependent
measurements indicate a classical origin. Furthermore,
Aharonov-Bohm type oscillations are observed at low temperatures
that correspond to enclosed areas much smaller than the size of a
lattice unit cell. These resonances are therefore attributed to
quasi-stable trajectories which are localized \emph{in between}
three neighboring antidots. Numerical simulations of the
magnetoresistivity based on the Kubo formalism support this
interpretation and allow us to specify the
characteristic trajectories.\\

The paper is organized as follows. In Section II, the sample
preparation and the experimental setup is discussed. Section III
is devoted to the experimental characterization of the resonances.
In Section IV, we present numerical simulations, identify the
relevant trajectories and relate these results to the experiments.
A summary and conclusion is given in Section V.

\section{\label{sec:2}SAMPLE PREPARATION AND EXPERIMENTAL SETUP}
Conventional modulation doped $GaAs/Al_{0.2}Ga_{0.8}As$ -
heterostructures with a two-dimensional electron gas (2DEG)
$83\rm{nm}$ below the surface have been grown by molecular beam
epitaxy. The 2DEG has an electron density of $n=2.5\times
10^{15}\rm{m^{-2}}$ and a carrier mobility of $90 \rm{m^2/Vs}$,
corresponding to an elastic mean free path of $7.6\rm{\mu m}$ and
to a Drude scattering time of $\tau= 35\rm{ps}$ at liquid helium
temperatures. A standard Hall bar geometry has been defined by
optical lithography and wet chemical etching. The 2DEG is accessed
electrically via $Ni/AuGe$ Ohmic contacts. The hexagonal antidot
arrays have been patterned by electron beam lithography and
subsequent ion beam etching with low-energy ($350\rm{eV}$)Ar$^+$
ions \cite{Lee1997}. The etch depth was $50\rm{nm}$. All antidot
arrays have a lattice constant of $600\rm{nm}$, while lithographic
antidot diameters of $d_l=100\rm{nm}$, $200\rm{nm}$, $250\rm{nm}$
and $320\rm{nm}$ have been patterned. The measured array area was
$50\rm{\mu m} \times 25\rm{\mu m}$.

Transport experiments have been carried out in a helium gas flow
cryostat with a variable temperature insert and a superconducting
magnet with a maximum field of $B = 8\rm{T}$, which is applied
perpendicular to the plane of the 2DEG. The temperature has been
varied from $1.6\rm{K}$ to $24\rm{K}$ with an accuracy of
$0.1\rm{K}$. For measurements at lower temperatures, a top loading
dilution refrigerator with a base temperature of $80\rm{mK}$ has
been used. An AC current (frequency $33\rm{Hz}$, amplitude
$5\rm{nA}$) was passed through the antidot array. The longitudinal
($\rho_{xx}$) and Hall ($\rho_{xy}$) components of the resistivity
tensor have been determined from voltage differences measured in
the corresponding four-probe setups.

\section{\label{sec:3}EXPERIMENTAL RESULTS}
\begin{figure}
\includegraphics[scale=0.5]{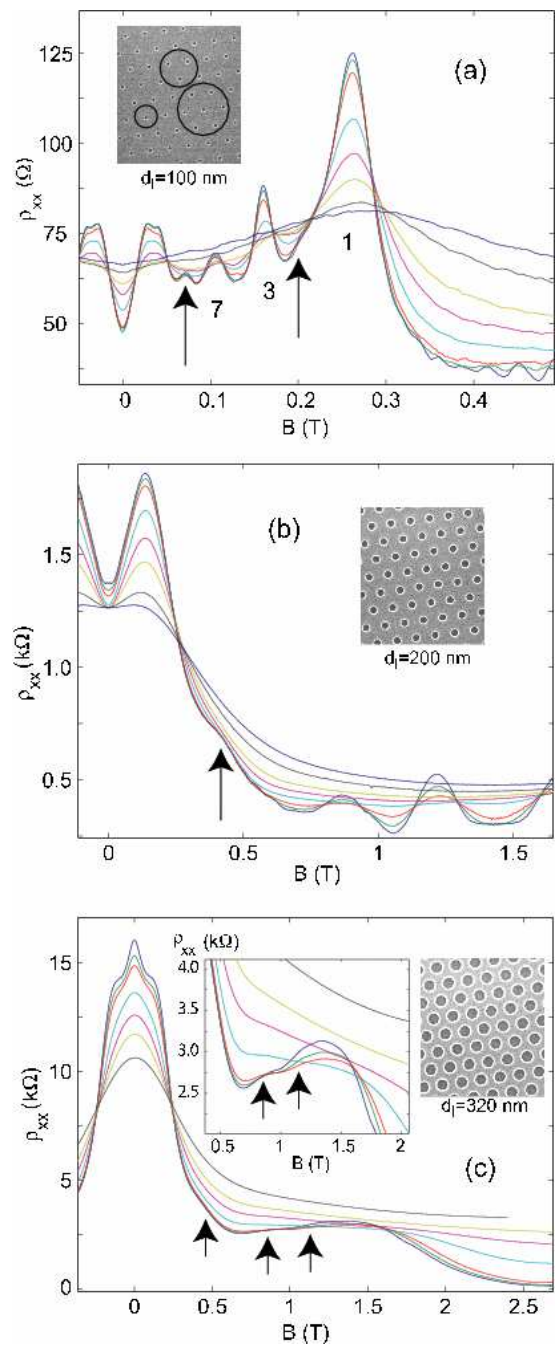}
\caption{\label{fig:1} Longitudinal magnetoresistivity of the
hexagonal antidot arrays at temperatures of $T=1.5\rm{K}$,
$2.5\rm{K}$, $4\rm{K}$, $8\rm{K}$, $12\rm{K}$, $16\rm{K}$, and
$24\rm{K}$, as observed for arrays with $a=600\rm{nm}$ and
different values of $d_{l}$ (inset). In (a), the well known
resonances corresponding to cyclotron orbits around 1, 3, and 7
antidots are observed. In (b) and (c), a known resonance is
observed around $B=130\rm{mT}$. The additional resonances are
indicated by arrows. The inset in (c) shows a zoom-in from the
main figure of the region of interest.}
\end{figure}
Fig. 1 shows the measurements of $\rho_{xx}$ as a function of $B$
for arrays with lithographic antidot diameters of
$d_l=100\rm{nm}$, $200\rm{nm}$ and $320\rm{nm}$. Due to lateral
depletion of the 2DEG around the antidots of about $100\rm{nm}$,
their electronic diameters $d_e$ are larger. They can be estimated
from the Shubnikov-de Haas oscillations observed in the antidot
arrays, which give the Fermi wavelength in the array as well as,
via their onset, the number of occupied transverse modes at the
bottleneck formed by adjacent antidots. This way, effective
electronic diameters of $d_e\approx 300\rm{nm}$, $400\rm{nm}$ and
$550\rm{nm}$, respectively, are found.

In the array with the smallest antidots (Fig. 1 (a)), typical
commensurability oscillations in $\rho_{xx}$ are observed that,
within the simple commensurability picture, in which the cyclotron
diameter $r_c=m^* v_F/(eB)$ matches the lattice constant in the
resistivity maxima, can be attributed to orbits around 1, 3, and 7
antidots \cite{Weiss1994}. Here, the resonance around one antidot
occurs at $B=270\rm{mT}$. In comparison to the decay of the
Shubnikov - de Haas oscillations with increasing temperature, the
temperature dependence of the commensurability oscillations is
weak, which indicates their classical origin. Note that there is
an additional shoulder at $B=200\rm{mT}$ and a peak at $70\rm{mT}$
with a weak temperature dependence (vertical arrows in Fig. 1(a)),
which cannot be identified in this commensurability picture.

Fig. 1 (b) shows the corresponding measurements on a sample with
$d_l=200\rm{nm}$. Here, only one strong resonance with a classical
character is visible at $B=130\rm{mT}$, which from simulations as
described below can be attributed to a superposition of the
commensurability oscillations identified in Fig. 1 (a). In
addition, a weak resonance is observed at $B=400\rm{mT}$. It is
clearly not related to the Shubnikov - de Haas oscillations that
set in around $B=0.6\rm{T}$, and shows a weak temperature
dependence. Such a structure is also present in an array with
$d_l= 250\rm{nm}$ (not shown). Moreover, in Fig.
 1 (b),  a small feature is observed between $B=0.5\rm{T}$ and $0.7\rm{T}$
which, however, cannot be clearly distinguished from Shubnikov -
de Haas oscillations. In arrays with even larger antidot diameters
(Fig. 1 (c)), three maxima, superimposed on a Shubnikov - de Haas
resonance, are detected at $B\approx 0.45\rm{T}$, $0.87\rm{T}$,
and $1.15\rm{T}$. The main focus of the present paper is on the
investigation of these resonances. The corresponding Hall
measurements are shown in Fig. 2. While the structure of the
conventional commensurability oscillations is reflected in
$\rho_{xy}(B)$, no signatures of the novel resonances are detected
here.
\begin{figure}
\includegraphics[scale=0.5]{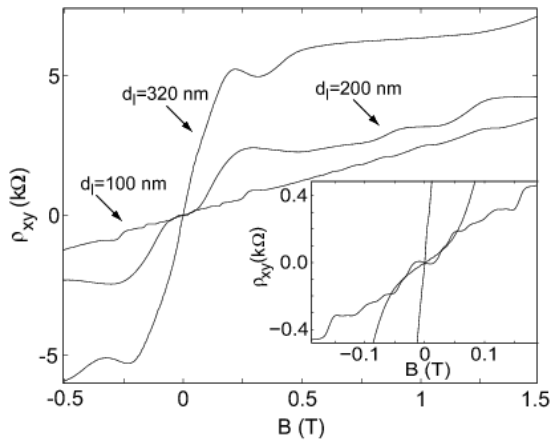}
\caption{\label{fig:2} Hall resistivities of the 3 arrays of Fig.
1, measured at $T=1.5\rm{K}$. In contrast to the conventional
commensurability oscillations, the additional resonances cannot be
identified. Note the negative Hall effect in the array with the
small antidots (inset). }
\end{figure}

\begin{figure}
\includegraphics[scale=0.5]{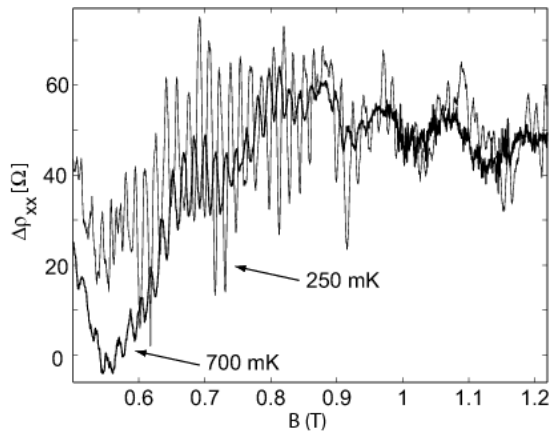}
\caption{\label{fig:3} Magnetoresistivity of the array with
$d_l=320\rm{nm}$ as measured in the dilution refrigerator in the
magnetic field regime of the novel resonances. A smooth background
has been subtracted. Here, the quasi-periodic Aharonov-Bohm
oscillations are modulated with a period of $\delta B\approx
120\rm{mT}$. }
\end{figure}

To obtain further insight into the character of these structures,
we have studied the magnetotransport of the array with
$d_l=320\rm{nm}$ in a dilution refrigerator, see Fig. 3. Around
$B=0$, the well-known Altshuler-Aronov-Spivak oscillations with a
period of $\delta B= 8\rm{mT}$ are observed (not shown)
\cite{Nihey1995}, which evolve into quasi-periodic Aharonov-Bohm
type oscillations with a period of $\delta B =16 \rm{mT}$ for
$B>60\rm{mT}$. The Aharonov-Bohm period corresponds to a
characteristic area of $A=h/e \delta B =2.6\times 10^{-13}m^{2}$,
in rough agreement with the area of a lattice unit cell
$(A_{cell}=3.1\times 10^{-13}m^{2})$. This evolution has been
reported and interpreted previously \cite{Ueki2004}. An
additional, superimposed oscillation, however, with a period of
$\delta B\approx 120\rm{mT}$ is observed in the magnetic field
regime where the novel classical resonances are detected.
Interpreting this oscillation in terms of an Aharonov-Bohm effect,
the enclosed area equals $A\approx 3.4\times 10^{-14}m^{2}$, which
corresponds to about $11\%$ of a unit cell. All periodic
oscillations smear out as the temperature is increased, with the
amplitude decaying roughly $\propto 1/T$; the oscillation with the
larger period persists to higher temperatures.

\section{\label{sec:4}INTERPRETATION AND DISCUSSION}

We interpret our results in terms of commensurability oscillations
that exist in the open electron pockets formed in between three
adjacent antidots. Quasi-stable orbits may form by consecutive
reflections of the electrons at the antidot walls. It seems
plausible that such trajectories become more stable as $d_e/a$
increases. We determine the characteristic area of such a pocket
from $A_{pocket}=\frac{1}{2}(A_{cell}-\pi(d_e/2)^2)$ and estimate
in our large antidot array to $A_{pocket}\approx 3.6\times
10^{-14}m^{2}$, i.e., significantly smaller than the area of the
unit cell. Hence, an increased Aharonov-Bohm period can be
expected, although the exact period depends on the details of the
closed trajectory.

In order to substantiate this interpretation, we have performed
model calculations based on the Kubo formalism \cite{Kubo1957}.
The antidot lattice is modelled by a two-dimensional, hexagonal
array of hard-wall cylinders of the estimated electronic diameter
$d_e$ for the samples of Fig. 1. For each value of $B$, $10^{5}$
electrons with a fixed Fermi energy adapted to the experimental
conditions ($E_F=8.86\rm{meV}$) are injected at random positions
and with random velocity directions within a unit cell of the
lattice. Their trajectories are calculated to a length of $50\mu
m$, corresponding to a time of flight of $240\rm{ps}$. The
electron velocity correlation function $\langle
v_i(t,B)v_j(0)\rangle$ is calculated, where the brackets denote
averaging over all trajectories \cite{Richter2000}. Within the
Kubo formalism, the components of the magneto-conductivity tensor
for a degenerate two-dimensional electron gas are obtained from

\begin{eqnarray}
\sigma_{ij}(B)=\frac{m^{*}e^{2}}{\pi\hbar^{2}}\int\limits_{0}^{\infty}
\langle v_i(t,B)v_j(0)\rangle e^{-t/\tau}dt \label{eq:one}
\end{eqnarray}

In Eq.~(\ref{eq:one}), $m^{*}=0.067 m_e$ is the effective electron
mass in GaAs. Elastic scattering due to random impurities is taken
into account via the exponential cutoff function that appears in
the integrand. Here, a Drude scattering time of $\tau=35\rm{ps}$
has been chosen, in accordance with the mobility of the pristine
2DEG. This model is a simplified version of the more thorough
treatment presented in Ref. \cite{Fleischmann1992}. In particular,
neither the driving electric field nor the finite slope of the
antidot walls are taken into account. Nevertheless, our simplified
model allows us to identify the trajectories that generate the
additional resonances. In Fig. 4, the results of the simulated
$\rho_{xx}(B)$ for the three arrays under study are reproduced.
For the largest antidots ($d_e=550\rm{nm}$), pronounced
resistivity minima are found around $B=0.78\rm{T}$, $1.4\rm{T}$,
and $1.85\rm{T}$, while the commensurability resonance around a
single antidot is barely visible at $B\approx 0.3\rm{T}$.

To gain more insight into the origin of these structures, we have
taken Poincar$\rm{\acute{e}}$ sections along the perpendicular
bisector of the line connecting the centers of two neighboring
antidots, see Fig. 5. As the magnetic field increases to
$B=0.6\rm{T}$, stable regions emerge with a structure as shown in
Fig. 5 (a). Around the resistivity maximum at $1\rm{T}$, this
pattern fades and evolves into a second one shown in Fig. 5(b),
which is most pronounced around the minimum in $\rho_{xx}$ at
$1.4\rm{T}$. As one goes to larger magnetic fields, the stable
regions gradually fade until they can no longer be identified. We
pick the Poincar$\rm{\acute{e}}$ sections at $B=0.6\rm{T}$ and
$1.2\rm{T}$ for a detailed discussion. In both cases, stable
regions are found which contain the two types of trajectories
shown in the right parts of Figs. 5 (a) and (b), respectively. The
first type is labelled by $\bf{1}$ and consists of skipping orbits
around an individual antidot. This type is present for all
magnetic fields above $0.5\rm{T}$ and covers a continuously
increasing area in the Poincar$\rm{\acute{e}}$ section as $B$ is
increased up to $2\rm{T}$. Such skipping orbit trajectories have
already been discussed for square antidot lattices
\cite{Eroms1999}. The second type are triangularly or rosette-type
shaped trajectories that form quasi-closed figures by consecutive
reflections at walls of adjacent antidots. They are labelled by
$\bf{2}$ - $\bf{4}$ in Fig. 5. The areas of the corresponding
regular islands in the Poincar$\rm{\acute{e}}$ sections have broad
peaks, centered around the minima in $\rho_{xx}$ at $B=0.78\rm{T}$
and $1.4\rm{T}$. Type 2 - trajectories with two consecutive
bounces at the same wall are present only for $B>1\rm{T}$.
\begin{figure}
\includegraphics[scale=0.5]{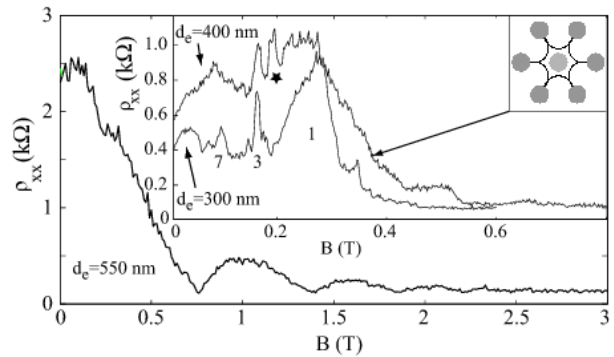}
\caption{\label{fig:4} Simulated longitudinal magnetoresistivity
of the array with $d_e=550\rm{nm}$. Weak resonances are found in
$0.6\rm{T}\leq B\leq 2.2\rm{T}$. The inset shows the corresponding
simulations for the arrays with $d_e=300 \rm{nm}$ and $d_e=400
\rm{nm}$. Here, the star denotes an unidentified resonances, while
the cartoon to the upper right sketches the type of trajectory
found in the feature indicated by the arrow. }
\end{figure}
Hence, we attribute the structure in the resistivity of the array
with the large antidots to the triangular or rosette - shaped
trajectories. They are most important around the minima of
$\rho_{xx}$, which is a consequence of the large magnetic field,
since localized states that generate a minimum in $\sigma_{xx}(B)$
translate into a minimum in $\rho_{xx}(B)$ for $\rho_{xy}\gg
\rho_{xx}$, as is the case here. Thus, a characteristic feature of
the novel commensurability resonances which distinguishes them
from those known previously is the fact that the magnetic field at
which the quasi-stable trajectories are most pronounced
corresponds to a minimum in $\rho_{xx}$.\\

Despite this qualitative agreement between the experimentally
observed features and the simulations, quantitative differences
are apparent. In particular, the positions of the measured minima
and maxima in $\rho_{xx}$ differ from the simulated ones, and
their amplitudes are weaker, an effect that, due to the observed
temperature independence, cannot be attributed to thermal
smearing. Also, the structure stemming from rosette-type
trajectories can be identified in simulations of the Hall
resistivity (not shown), in contrast to our Hall measurements. One
possible reason for these discrepancies may be non-specular
scattering at the antidot walls. Low-energy ion beam milling has
been the technique of choice for patterning of the antidots, since
it is established as a method to generate nanostructures with
walls of high specularity around 85\% \cite{Lee1997}. We expect
that the small fraction of non-specular scattering may lead to
some degree of smearing, but most likely not to a displacement of
the extremal points. Rather, we interpret these deviations as a
manifestation of soft antidot walls, probably in combination with
some disorder in the antidot sizes, shapes and positions, as well
as the driving electric field. We are not aware of experimental
work that determines the wall steepness for the etching technique
we used. However, it seems plausible to assume that the steepness
is comparable to that one obtained in Ga[Al]As samples of similar
electron density, patterned by other dry etching techniques, like
those used in Ref. \cite{Weiss1991}. More elaborate simulations
that use the wall steepness as a parameter \cite{Fleischmann1992,
Fleischmann1994} have been able to establish a close to perfect
agreement with the experimental observations of Ref.
\cite{Weiss1991}. As an example, we note that a peak is observed
in the experiments on antidot arrays with small diameters at small
magnetic fields (Fig. 1 (a)), which cannot be attributed to a
trajectory by our simulations. A similar discrepancy in square
lattices \cite{Weiss1991} could be resolved by the softness of the
walls \cite{Fleischmann1992}, a fact which suggests that the
deviations in our study can be explained by soft walls as well. In
general, soft walls tend to deform the trajectories, which in turn
causes some resonances to be significantly displaced, while the
sensitivity to the wall steepness depends on the resonance
\cite{Fleischmann1992}. Such a deformation would be in tune with
the observation that the areas enclosed by the simulated
trajectories (Fig. 5) are smaller than the measured enclosed areas
(Fig. 3). Also, the peak around $0.35\rm{T}$ in the simulated
array with $d_e=300\rm{nm}$ can be identified in a
Poincar$\rm{\acute{e}}$ section of the hard wall simulation (Inset
in Fig. 4), but is not observed experimentally. It seems plausible
to us that such a trajectory is very sensitive to an even moderate
softening of the wall. To study whether more realistic
simulations, or experiments on samples with steeper walls
\cite{Eroms1999}, are able to resolve the discrepancies,
is beyond the present study. \\
For the sample with the smallest antidot diameter, the simulations
reproduce the experiments remarkably well in the range $r_c\geq a$
(Figs 1 (a), 2 and 4). The measured negative Hall (Fig. 2) effect
is reproduced in our hard wall simulations (not shown), in
contrast to hard wall simulations performed for square lattices
\cite{Fleischmann1994}. In square lattices consisting of antidots
with similar ratios $d/a$, simulations do not suggest the
existence of type 2 - orbits, independent of the wall steepness
\cite{Eroms1999}. Thus, the wall steepness seems less critical in
hexagonal arrays, but becomes more relevant as $d_e/a$ increases.
We therefore conclude that a hexagonal lattice strongly favors the
formation of stable orbits in between antidots as compared to
square lattices.

\begin{figure}
\includegraphics[scale=0.5]{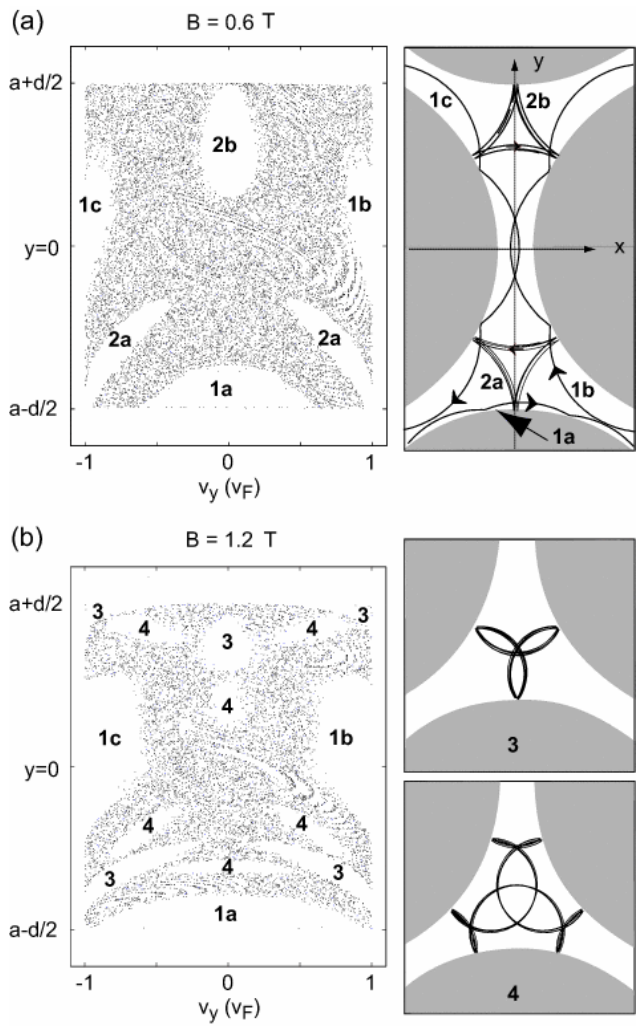}
\caption{\label{fig:5} Poincar$\rm{\acute{e}}$ - sections (left)
taken from the array with $d_e=550\rm{nm}$ at $B=0.6\rm{T}$ (a)
and $1.2\rm{T}$ (b). The sections are taken along the
$y$-direction at $x=0$, see the right part of (a), where the gray
areas denote the antidots. The characteristic trajectories (right)
found in the stable regions are assigned according to the numbers
in the stable regions, see text. Note that only the intersections
of the trajectories with $v_x>0$ are plotted in the
Poincar$\rm{\acute{e}}$ - sections. }
\end{figure}

\section{\label{sec:5}SUMMARY AND CONCLUSION}

In summary, we observe novel resonances in hexagonal antidot
lattices with large antidot diameters. Classical simulations
suggest that they can be attributed to classical, quasi-periodic
trajectories of various shapes that are caught in between three
adjacent antidots. This interpretation is supported by
measurements at temperatures below $1\rm{K}$, where the samples
show Aharonov-Bohm type oscillations with a characteristic area
similar to that one of an electron pocket in between antidots.  It
remains to be seen whether such resonances are also visible in
other types of antidot lattices with optimized parameters. In
order to improve the quantitative agreement with simulations, it
would be interesting to perform more sophisticated calculations
with realistic potential landscapes, inclusion of the driving
electric field, disorder, and the wave nature of the electrons.\\

Financial support by the \emph{Heinrich-Heine-Universit\"at
D\"usseldorf} and by the \emph {Region Ile de France} and the
\emph{Conseil G$\rm{\acute{e}}$n$\rm{\acute{e}}$ral de l'Essonne}
is gratefully acknowledged. The authors acknowledge technical
support by Ch. Dupuis.

\newpage
\bibliography{Meckler}% Produces the bibliography via BibTeX.

\end{document}